\documentclass[aps, nofootinbib]{revtex4}
\usepackage{psfrag}
\usepackage{array}
\usepackage{amstext,amsmath,amssymb,amsfonts,bbm}
\usepackage[dvips]{graphicx}
\usepackage{epsfig}
\usepackage[utf8]{inputenc}

\newcommand{\N}{\mathbb{N}}

\newcommand{\f}{\frac}

\newcommand{\pd}{\partial}

\newcommand{\SU}{\mathrm{SU}}

\def\la{\langle}
\def\ra{\rangle}

\newcommand{\cO}{{\cal O}}
\newcommand{\tcO}{\widetilde{{\cal O}}}

\def\eps{\epsilon}

\def\arr{\rightarrow}

\def\nn{\nonumber}

\newcommand{\be}{\begin{equation}}
\newcommand{\ee}{\end{equation}}
\newcommand{\bes}{\begin{eqnarray}}
\newcommand{\ees}{\end{eqnarray}}

\newcommand{\Ref}[1]{(\ref{#1})}

\newcommand{\sixj}[2]{\left\{\begin{array}{ccc} #1 \\ #2 \end{array}\right\}}
\def\sj{{\{6j\}}}

\def\vtheta{\vartheta}

\newcommand{\equa} [1] {\begin{equation} #1\end{equation}}

\newcommand{\tabl} [2] {\begin{array} {#1} #2 \end{array}}

\begin{document}

\title{\large\bf The 6j-symbol: Recursion, Correlations and Asymptotics}

\author{Ma\" \i t\'e Dupuis}\email{maite.dupuis@ens-lyon.fr}
%
\author{Etera R. Livine}\email{etera.livine@ens-lyon.fr}
\affiliation{Laboratoire de Physique, ENS Lyon, CNRS-UMR 5672, 46 All\'ee d'Italie, Lyon 69007, France}

\date{\small \today}

\begin{abstract}

We study the asymptotic expansion of the \sj-symbol using the Schulten-Gordon recursion relations. We focus on the particular case of the isosceles tetrahedron and we provide explicit formulas for up to the third order corrections beyond the leading order. Moreover, in the framework of spinfoam models for 3d quantum gravity, we show how these recursion relations can be used to derive Ward-Takahashi-like identities between the expectation values of graviton-like spinfoam correlations.

\end{abstract}

\maketitle

\section{Introduction: \sj-symbol and the recursion formula}
\label{intro}

Spinfoam models present a proposal for a well-defined path integral formulation for quantum gravity. They define the structure quantum space-time at the Planck scale. The current challenge is to study their semi-classical behavior at large scale, show that we recover general relativity and a locally flat space-time as expected and then to extract the perturbative quantum gravity correction to the classical gravitational dynamics. The main spinfoam models for 4d quantum gravity are the Barrett-Crane model \cite{BC} and the more recent family of models \cite{EPR,FK,LS} based on coherent state techniques. These models are all derived from the reformulation of general relativity as a constrained topological BF-theory and attempt to define a discretization of the path integral over space-time geometries.

A recent proposal, which is actively investigated, to probe the semi-classical behavior of the spinfoam amplitudes is the test of the graviton propagator proposed by Rovelli and collaborators \cite{graviton}. There have been several more or less rigorous analytical computations \cite{anal}, as well as numerical simulations \cite{numerics}, showing that the most recent spinfoam models have the expected behavior reproducing at first order (and for the simplest space-time triangulation) Newton's law and the tensorial structure of the graviton. These rely on the computation of the leading order asymptotics of some spinfoam amplitudes for large spin. The next step is to go beyond the leading order and compute the quantum corrections which should correspond to the loop corrections of the standard perturbative quantum field theory approach. This necessarily requires a better understanding of the structure of the asymptotics of the spinfoam amplitudes. Here, we investigate this issue in the simplified framework of 3d quantum gravity.

In three space-time dimensions, gravity is a topological theory which can be exactly quantized as a spinfoam model, the well-known Ponzano-Regge model \cite{PR}. Its basic building block is the \sj-symbol. Its asymptotics has been well-studied at leading order and been derived using various techniques: from the brute-force calculation based on the explicit formula of the \sj-symbol in term of factorials \cite{razvan} to more refined saddle point approximation on integral formulas \cite{6jsaddle}. Recently, these calculations have been pushed further in order to compute the corrections to the asymptotic behavior using both the saddle point technique \cite{valentin} and the brute-force method \cite{maite}. These results apply to the computation of graviton-like correlations in 3d quantum gravity \cite{3dtoymodel, valentin}. This should lead to a better understanding of the structure of the quantum corrections of the spinfoam graviton propagator, which should be relevant to the four-dimensional case.

In the present paper, we are interested in the computation of the asymptotics of the \sj-symbol through the use of recursion relations. As was first shown in \cite{SG}, the \sj-symbol satisfies a recursion formula which are intimately related to the topological invariance of the Ponzano-Regge spinfoam model. This formula turned out to be very useful: it can be approximated in the large spin limit by a second order differential equation and one can use it to derive the leading order of the asymptotics through a WKB approximation, but it also allows fast numerical calculations of the \sj-symbol.

Here, we investigate two aspects of these recursion relations.
First, we show how to extract the next-to-leading and subsequent corrections to the \sj-symbol. We focus on the case of the isosceles tetrahedron (since this is the case relevant for the computation of graviton-like correlations \cite{3dtoymodel}) and we compare our results with the previous work \cite{valentin, maite}.
Second, we study the consequences of the existence of such a recursion relation on the behavior of the graviton-like correlation functions in spinfoam models. We show that it leads to relations on these correlations functions, which relate the expectation values of different observables. These relations are similar to the Ward-Takahashi identities (and to the Schwinger-Dyson equation) in standard quantum field theory, which allow to relate different correlation functions of different orders. These equations play a key role in the study of the renormalization of quantum field theories, so we expect these new recursion relation to be equally relevant to the study of the renormalization/coarse-graining of spinfoam models.

\section{The Recursion Relation for the Isosceles Tetrahedron}

\subsection{Exact and Approximate Recursion}

We will focus on the isosceles tetrahedron, which is relevant for the computations of geometrical correlations in the simplest non-trivial toy model in 3d quantum gravity \cite{3dtoymodel}. Such a tetrahedron has four of its edges of equal length with the two remaining opposite edges of arbitrary length. The corresponding isosceles \sj-symbol is:
$$
\{a,b\}_J\equiv\sixj{a&J&J}{b&J&J},
$$
where $J\in \N/2$ and $a,b$ are integers smaller than $2J$ (to satisfy the triangular inequality). The associated tetrahedron has edge lengths $l_j=d_j/2$ for $j=a,b,J$, where $d_j=2j+1$ is the dimension of the $\SU(2)$ representation of spin $j$.
The volume $V$ of the tetrahedron is given by the simple formula:
\be \label{volume}
V_J(a,b)=\f1{12}l_al_b\sqrt{4l_J^2-\left(l_a^2+l_b^2\right)},
\ee
while the (exterior) dihedral angles $\theta_j$  can also be written in term of the edge lengths (see e.g. \cite{valentin} for more details):
\be \label{angles}
\cos\theta_a\,=\,
-\f{4l_J^2-l_a^2-2l_b^2}{4l_J^2-l_a^2},\quad
\cos\theta_b\,=\,
-\f{4l_J^2-l_b^2-2l_a^2}{4l_J^2-l_b^2},\quad
\cos\theta_J\,=\,
\f{-l_al_b}{\sqrt{4l_J^2-l_a^2}\sqrt{4l_J^2-l_b^2}}.
\ee
The general recursion relation for the \sj-symbol given by Schulten and Gordon in \cite{SG} simplifies in this specific isosceles case:
\be \label{exactrecursion}
(l_a+\frac{1}{2})\left[4l_J^2-(l_a+\frac{1}{2})^2\right]\{a+1,b\}_J
-2l_a\left[(4l_J^2-l_a^2)\cos(\theta_a)+\frac{1}{4}\right]\{a,b\}_J
+(l_a-\frac{1}{2})\left[4l_J^2-(l_a-\frac{1}{2})^2\right]\{a-1,b\}_J =0
\ee
In the asymptotic regime, we know (analytically and numerically) the behavior of the \sj-symbol at the leading order:
\be
\label{LO}
\{a,b\}_J \,\sim\,
\{a,b\}_J^{LO}\,\equiv
\f1{\sqrt{12\pi V}}\,\cos\left(
l_a\theta_a +l_b\theta_b +4l_J\theta_J +\f\pi4
\right),
\ee
which is actually valid under the assumption that the tetrahedron with edge lengths $l_a,l_b,l_J$ exists (else the generic asymptotics can be expressed in term of Airy functions). The oscillatory phase is given by the Regge action $S_R=l_a\theta_a +l_b\theta_b +4l_J\theta_J$.
Using the obvious trigonometric identity $\cos((n+1)\phi)+\cos((n-1)\phi)=\,2\cos\phi\,\cos n\phi$, we can write an exact recursion relation for the leading order of the \sj-symbol:
\be \label{exactrecurLO}
\sqrt{V_J(a+1,b)}\,\{a+1,b\}_J^{LO}
-2\cos\theta_a \,\sqrt{V_J(a,b)}\,\{a,b\}_J^{LO}
+\sqrt{V_J(a-1,b)}\,\{a-1,b\}_J^{LO}=0.
\ee
A similar recursion relation holds for $b$-shifts and also $J$-shifts.

The most natural idea is to compare this recursion relation for the leading order to the previous  equation on the exact \sj-symbol to see how to use them to extract the next-to-leading correction to the asymptotic behavior. We can first  find the link between the leading order of equation (\ref{exactrecursion}) and the leading order of equation (\ref{exactrecurLO}). Both equations can be written under the same form at the leading order:
\equa{
\{a+1,b\}_J -2\cos \theta_a \{a,b\}_J +\{a-1,b\}_J \approx 0,
}
which turns into a simple second order differential equation in the large spin limit.
Then the next-to-leading order of the equation (\ref{exactrecursion}):
\equa{\tabl{ll}{
\sqrt{V_J(a,b)}\left(1+ \f{1}{2l_a} \left(1-\f{2l_a^2}{4l_J^2-l_a^2}\right)\right)&\{a+1,b\}_J-2\cos \theta_a \sqrt{V_J(a,b)}\{a,b\}_J \\
&+\sqrt{V_J(a,b)}\left(1- \f{1}{2l_a} (1-\f{2l_a^2}{4l_J^2-l_a^2})\right)\{a-1,b\}_J \approx 0\\
}}
will have to be compared to an recursion relation for the next-to-leading order of the \sj-symbol.

\subsection{Pushing to the Next-to-Leading Order}

We are interested in the asymptotic expansion  of the \sj-symbol. It was shown in previous works \cite{valentin,maite} that  $l_j$ seems to be the right parameter to consider when studying the semi-classical behavior of the \sj-symbol. So from now we write:
$$
\sixj{a&J&J}{b&J&J}\equiv \{l_a, l_b\}_{l_J}.
$$
Notice that shifting $a$ by $\pm 1$ is equivalent to shifting the edge length $l_a=a+1/2$ by $\pm 1$. We rescale now $l_j$ by $\lambda l_j$ and we replace the exact \sj-symbol by a series in $1/\lambda$ alternating cosines and sinus of the Regge action (shifted by $\pi/4$) in the previous equation (\ref{exactrecursion}). The fact that there is no mixing up of cosines and sinus at all order was show in \cite{valentin}. More precisely, we write the \sj-symbol asymptotic expansion under the form:
%
\equa{\label{6jNNNLO}
\tabl{ll}{
\{\lambda l_a, \lambda l_b\}_{\lambda l_J}=\frac{1}{\lambda^{3/2}D(l_a,l_b,l_J)}[\cos(\lambda S_R+\pi/4)+ &\frac{F^{(1)}(l_a,l_b,l_J)}{\lambda}\sin(\lambda S_R+\pi/4)+\frac{G^{(1)}(l_a,l_b,l_J)}{\lambda}\cos(\lambda S_R+\pi/4))\\
&+ \frac{F^{(2)}(l_a,l_b,l_J)}{\lambda^2}\cos(\lambda S_R+\pi/4)+\frac{G^{(2)}(l_a,l_b,l_J)}{\lambda^2}\sin(\lambda S_R+\pi/4))\\
&+ \frac{F^{(3)}(l_a,l_b,l_J)}{\lambda^3}\sin(\lambda S_R+\pi/4)+\frac{G^{(3)}(l_a,l_b,l_J)}{\lambda^3}\cos(S_R+\pi/4)\\
&+ \frac{F^{(4)}(l_a,l_b,l_J)}{\lambda^4}\cos(\lambda S_R+\pi/4)+\frac{G^{(4)}(l_a,l_b,l_J)}{\lambda^4}\sin(S_R+\pi/4) + O(\lambda^{-5})],
}}
where the pre-factor denominator $D(l_a,l_b,l_J)$ is given by the square-root of the tetrahedron volume as in equation \Ref{LO}.
To study the asymptotics, it is convenient to factorize the whole equation (\ref{exactrecursion}) by $\lambda^{3/2}$. We then write $\{ l_a \pm 1/ \lambda, l_b \}_{l_J}$ for $\{\lambda l_a \pm 1, \lambda l_b \}_{\lambda l_J}$. We also factorize the coefficients of the recursion relation. We start by defining $C(l_j)=l_a(4l_J^2-l_a^2)=\frac{16(A(la,l_J,l_J))^2}{l_a}$ where $A(a,b,c)=\frac{1}{4}\sqrt{(a+b+c)(a+b-c)(a-b+c)(-a+b+c)}$ is the area of the triangle of edge lengths given by $a$, $b$ and $c$.
The coefficient which appears in front of $\{l_a \pm 1/\lambda, l_b\}_{l_J}$ becomes $C(l_a\pm1/(2\lambda),l_b,l_J)=(l_a\pm 1/(2\lambda))(4l_J^2-(l_a\pm1/(2\lambda))$, where we underline that the shift is $\pm1/(2\lambda)$ and not simply $\pm1/\lambda$.  We expand $C(l_a\pm1/(2\lambda),l_b,l_J)$ in term of derivatives:
$$
C(l_a\pm1/(2\lambda),l_b,l_J)= \sum_n \frac{1}{n!}\frac{1}{(2\lambda)^n}\frac{\partial^nC}{\partial l_a^n}
$$
with \equa{\label{coefequa}
\left\{
\tabl{l}{C=l_a(4l_J^2-l_a^2) \\
\f{\pd C}{\pd l_a}=4l_J^2-3l_a^2\\
\f{ \pd^2 C}{\pd l_a^2}=-6l_a^2\\
\f{ \pd^3 C}{\pd l_a^3}=-6\\
\f{\pd^nC}{\pd l_a^n}=0 \textrm{ for } n\geq 4}
\right.
}
Then to express  $\{ l_a \pm 1/\lambda, l_b\}_{ l_J}$ we need to expand $D(l_a\pm 1/\lambda)$, $F^{(i)}(l_a\pm 1/\lambda)$, and $G^{(i)}(l_a\pm 1/\lambda)$: $(i\in \{1\cdots 4\})$
 \equa{\label{coef6j}
 \left\{
 \tabl{l}{
D(l_a\pm 1/\lambda)= D\pm \f{1}{\lambda} \frac{\partial D}{\pd l_a}+\f{1}{2\lambda^2}\f{\pd^2 D}{\pd l_a^2}\pm \f{1}{3!\lambda^3}\f{\pd^3D}{\pd l_a^3} +\f{1}{4!\lambda^4}\f{\pd^4D}{\pd l_a^4}\\
F^{(i)}(l_a\pm 1/\lambda)=\sum_{k=0}^{4-i}(-1)^{k}\f{1}{k!\lambda^k}\f{\pd^k F^{(i)}}{\pd l_a^k} \; \\
G^{(i)}(l_a\pm 1/\lambda)=\sum_{k=0}^{4-i}(-1)^{k}\f{1}{k!\lambda^k}\f{\pd^k G^{(i)}}{\pd l_a^k} \\
}
\right.}
$F^{(1)}(l_j)$ was computed in a previous paper \cite{valentin,maite}. It was also suggested that the asymptotic expansion of the \sj-symbol in term of the length  scale $\lambda$ is given by an alternative of cosines and sinus at each order, so we expect that $G^{(i)}(l_j)=0$ for $\forall i\ge 1$. Finally, we also need to expand the Regge action $\lambda S_R(l_a\pm \frac{1}{\lambda})$, remembering that $\theta_j=\theta_j(l_a)$~:
 \equa{ \label{action}
\lambda S_R(l_a\pm \frac{1}{\lambda})= \lambda S_R +\displaystyle{\sum_{k=0}^{4}}\f{(-1)^{k+1}}{(k+1)!\lambda^k}\f{\pd^k \theta_a}{\pd l_a^k}
}
 with $$\left\{ \tabl{l}{ \f {\partial \theta_a}{\partial l_a}= \f{-2l_a l_b}{(4l_J^2-l_a^2)\sqrt{4l_J^2-l_a^2-l_b^2}} \\
\\
\f{\pd^2\theta_a}{\pd l_a^2}= - \f{2l_b(4l_J^2l_a^2-2l_a^4-l_a^2l_b^2+16l_J^4-4l_b^2l_J^2)}{(4l_J^2-l_a^2)^2[4l_J^2-l_a^2-l_b^2]^{3/2}}\\
\\
\f{\pd^3 \theta_a}{\pd l_a^3}=-\f{2l_al_b(24l_b^4l_J^2+40l_J^2l_a^2l_b^2-12l_J^2l_a^4+5l_a^4l_b^2-192l_J^4l_a^2-240l_J^4l_b^2+2l_a^2l_b^4+6l_a^6+576l_J^6)}{(4l_J^2-l_a^2)^3[4l_J^2-l_a^2-l_b^2]^{5/2}}  \\
\\
\f{\pd^4 \theta_a}{\pd l_a^4}= \f{1}{(4l_J^2-la^2)^4(4l_J^2-l_a^2-l_b^2)^{7/2}}(6(8l_a^{10}+8l_a^8l_b^2+152l_J^2l_a^6l_b^2-720l_J^4l_a^6+7l_a^6l_b^4+3520l_J^6l_a^4-1472l_J^4l_a^4l_b^2+2l_a^4l_b^6\\
\quad \quad \quad+140l_a^4l_b^4l_J^2-560l_a^2l_b^4l_J^4-3840l_J^8l_a^2+2432l_J^6l_a^2l_b^2+48l_a^2l_J^2l_b^6+2048l_J^8l_b^2-448l_J^6l_b^4-3072l_J^{10}+32l_b^6l_J^4)l_b)
}\right. $$
We can now write an asymptotic recursion equation from equations (\ref{exactrecursion}), (\ref{coefequa}), (\ref{6jNNNLO}), (\ref{coef6j}) and (\ref{action}) in terms of $\lambda$ neglecting terms of order $O(\lambda^{-4})$ and smaller, assuming that $\lambda$ is large. This leads to a couple of equations at each order, one for the $\cos$-oscillations and one for the term in $\sin$:
\begin{itemize}
\item The first equation is given by the terms of order $\lambda^{0}$ and it is trivially satisfied $(0=0)$ since we have already written  the leading order of the \sj-symbol proportional to $\cos(S_R+\f\pi4)$ (the Ponzano-Regge asymptotic formulae).
\item The second equation is given by the terms of order $\lambda^{-1}$:
\equa{ \label{equa0}
\left( \f{1}{2C}\f{\pd C}{\pd l_a}-\f1D\f{\pd D}{\pd l_a}\right)\sin(\theta_a) +\f12 \f{\pd \theta_a}{\pd l_a} \cos(\theta_a)=0
}
which can be rewritten as a differential equation for $D$:
\equa{ \label{equa1}
\frac{\partial \ln D}{\partial l_a}=\frac{1}{2} \left[ \frac{\partial \theta_a}{\partial l_a} \frac{\cos \theta_a }{\sin \theta_a} + \frac{\partial \ln C}{\partial l_a} \right].
}
This allows to determine $D$: $\ln D= \f12 \ln(C\sin(\theta_a))+ K$, which simplifies into $D=K\sqrt{ l_a l_b \sqrt{4l_J^2-l_a^2-l_b^2}}$ where $K$ is a constant factor. Thus this second equation shows that $D$ is correctly proportional to the square-root of the volume $V$ of the isosceles tetrahedron. To determine the normalization constant $K$ (as well as $G^{(1)}$), the orthonormality property of \sj-coefficients can be employed: $\sum_a 4l_a \sqrt{l_bl_{b^\prime}} \{a,b\}_J \{a,b^\prime \}_J= \delta_{b b^\prime }$ and we get the $K=\sqrt{12\pi}$. The details are given in the next section.

\item The third equation is given by the terms of order $\lambda^{-2}$ and which are proportional to $\cos (S_R + \f\pi4)$
\equa{ \tabl{ll}{ \label{equa2}
 \f{\pd F^{(1)}}{\pd l_a}=\f{l_a}{4C \sin \theta_a} + &\left( \f{1}{2C}\f{\pd C}{\pd l_a } -\f1D\f{\pd D}{\pd l_a}\right) \f12 \f{\pd \theta_a}{\pd l_a}+\f16 \f{\pd^2 \theta_a}{\pd l_a^2}\\
 &+ \f{\cos \theta_a}{\sin \theta_a} \left(\f{1}{2D}\f{\pd^2 D}{\pd l_a^2}-\f{1}{8C}\f{\pd^2 C}{\pd l_a^2}+\f1D\f{\pd D}{\pd l_a}\left(\f{1}{2C}\f{\pd C}{\pd l_a}-\f1D \f{\pd D}{\pd l_a}+(\f12\f{\pd \theta_a}{\pd l_a})^2 \right) \right)
}}
where we used the fact that $\left(\f{1}{2C}\f{\pd C}{\pd l_a}-\f1D\f{\pd D}{\pd l_a}\right)\sin(\theta_a) +\f12 \f{\pd \theta_a}{\pd l_a} \cos(\theta_a)=0$ (eqn. \Ref{equa1}) to remove all the terms proportional to $F^{(1)}$ itself.
The first term of the right-hand side of the equation (\ref{equa2}) comes from the variation of the coefficient in front of $\{a,b\}_J$ in the recursion equation (\ref{exactrecursion}). The terms with a derivative of $C$ with respect to $l_a$ come from the coefficients in front of $\{a\pm 1, b\}_J$ and $\{a,b\}_J$. The variation of $C$ with respect to $l_a$ is given by the variation of the areas of the triangles of the tetrahedron. From eqn.(\ref{equa1}), we relate it to the variations of $D$ (the volume) and to the variations of the dihedral angle $\theta_a$: $\f1C \f{\pd C}{\pd l_a}=\f2D \f{\pd D}{\pd l_a} -\f{\cos \theta_a}{\sin \theta_a}\f{\pd \theta_a}{\pd l_a}$. The terms with a derivative of $D$ with respect to $l_a$ come from the variation of the leading order of the asymptotic of the \sj-symbol and the terms with a derivative of the dihedral angle $\theta_a$ come from the variations of the Regge action $S_R$. We can now compute the derivative of $F^{(1)}$ with respect of $l_a$ (equation (\ref{equa2})) in terms of $l_a$, $l_b$ and $l_J$ the edge lengths of the tetrahedron:
\equa{\tabl{ll}{
 \f{\pd F^{(1)}}{\pd l_a}=-&\f{1}{48(l_a^2(-4l_J^2+l_a^2)^2(4l_J^2-l_a^2-l_b^2)^{(5/2)}l_b)}(-32l_b^6l_J^2l_a^2+10l_b^6l_a^4+96l_b^6l_J^4-960l_J^6l_b^4+15l_a^6l_b^4+400l_J^4l_b^4l_a^2-100l_J^2l_b^4l_a^4\\
 &-168l_J^2l_a^6l_b^2-1664l_J^6l_b^2l_a^2+20l_a^8l_b^2+576l_J^4l_b^2l_a^4+3072l_J^8l_b^2-3072l_J^{10}+48l_J^4l_a^6+2304l_J^8l_a^2-576l_J^6l_a^4)
 }}
 and then easily integrate this equation over $l_a$:
\equa{\tabl{ll}{
F^{(1)}(l_j)&=\\
&-\frac {768{l}_J^{6}(l_J^2-l_a^2-l_b^2)+736{l}_J^{4}l_a^{2}l_b^{2}+240l_J
^{4}(l_a^{4}+l_b^{4})-176{l}_J^{2} l_a^
{2}l_b^{2}(l_a^2+l_b^2)-24{l}_J^{2}( l_a^{6}+l_b^{6})+10l_a^{2} l_b^
{2}(l_a^4+l_b^4)+25 l_a^{4}l_b^{4}}
{ 24\left( 4l_J^2-l^2_b \right)\left( 2l_J^2-l_a^2 \right)\left( 4{l}_J^{2}-l_b^{
2}-l_a^{2} \right)^{3/2}l_al_b }  +Z(l_b, l_J)
}}
The integration constant $Z(l_b, l_J)$ can be determined using the symmetry properties of the \sj-symbol: symmetry of the isosceles \sj-symbol with respect to $l_a$ and $l_b$, coupling of $l_a$, $l_b$ and $l_J$ by this isosceles \sj-symbol and homogeneity of $F^{(1)}$ ($[F^{(1)}]=l_j^{-1}$) imply that $Z(l_b,l_J)=0$. Then this gives us the same result as in the previous paper \cite{maite}. Moreover, using the definitions of the tetrahedron volume (\ref{volume}) and of the dihedral angles (\ref{angles}), we can express $F^{(1)}$ in terms of some geometrical characteristics of the tetrahedron:
\equa{ \label{geometricNLO}
F^{(1)}=-\f{\cos \theta_J\left(3(12V)^8-(12V)^4l_a^4l_b^4\left(3(l_a^2-l_b)^2+2l_a^2l_b^2\right)-l_a^{12}l_b^{12} \right)+6l_a^{12}l_b^{12}}{48(12V)^3l_a^8l_b^8}
}
\item The fourth equation is given by the terms of order $\lambda^{-2}$ and which are proportional to $\sin (S_R + \f\pi4)$. It is the same equation as the previous one for $G^{(1)}$  but the right-hand side is now equal to zero (homogenous equation). That is we simply get that
\equa{
 \f{\pd G^{(1)}}{\pd l_a}= 0
}
so $G^{(1)}=Z(l_b,l_J)$ is just a constant of integration. Once again the symmetry properties of the \sj-symbol implies that $G^{(1)}=0$.
\item The next equation is given by the terms of order $\lambda^{-3}$ and which are proportional to $\sin (S_R + \f\pi4)$. We get an equation for the first derivative of $F^{(2)}$ with respect to $l_a$
\equa{ \tabl{l}{\label{equa5}
\f{\pd F^{(2)}(l_j)}{\pd l_a}= \f{\cos \theta_a}{2 \sin \theta_a} \f{\pd^2F^{(1)}}{\pd l_a^2} - \left(\f{1}{\sin^2\theta_a}\f{\pd \theta_a}{\pd l_a} + F^{(1)}\right) \f{\pd F^{(1)}}{\pd l_a} \\
\;\;\;+ \f{\cos \theta_a}{\sin \theta_a} \left[  -\f{1}{4!}\f{\pd^3 \theta_a}{\pd l_a^3}+\left(\f1D \f{\pd D}{\pd l_a}\f{1}{2C}\f{\pd C}{\pd l_a} + \f{1}{2D}\f{\pd^2D}{\pd l_a^2}-\left(\f1D\f{\pd D}{\pd l_a}\right)^2-\f{1}{8C}\f{\pd^2C}{\pd l_a^2}+ \f{1}{3!}\left(\f12\f{\pd \theta_a}{\pd l_a}\right)^2\right)\f12 \f{\pd \theta_a}{\pd l_a} +\left(\f1D\f{\pd D}{\pd l_a}-\f{1}{2C}\f{\pd C}{\pd l_a}\right) \f{1}{3!}\f{\pd^2\theta_a}{\pd l_a^2}  \right] \\
\;\;\; +\f{1}{2C}\f{\pd C}{\pd l_a}\f{1}{2D}\f{\pd^2 D}{\pd l^2_a}+\f{1}{8C}\f{\pd^2 C}{\pd l^2_a}\f{1}{D}\f{\pd D}{\pd l_a} +\left( \f{1}{2C}\f{\pd C}{\pd l_a}-\f{1}{D}\f{\pd D}{\pd l_a}\right) \f12 \left(\f{1}{2}\f{\pd \theta_a}{\pd l_a}\right)^2 +\f{1}{3!}\f{\pd^2 \theta_a}{\pd l^2_a}\f{1}{2}\f{\pd \theta_a}{\pd l_a} +\left(\f{1}{D}\f{\pd D}{\pd l_a}\right)^3-\f{1}{D}\f{\pd D}{\pd l_a}\f{1}{D}\f{\pd^2 D}{\pd l^2_a}+ \f{1}{3!D}\f{\pd^3 D}{\pd l^3_a}\\
\;\;\; -\f{1}{8\cdot 3! C}\f{\pd^3 C}{\pd l^3_a} -\f{1}{2C}\f{\pd C}{\pd l_a}\left(\f{1}{D}\f{\pd D}{\pd l_a}\right)^2
}}
We recall that $D$ is proportional to the square root of the tetrahedron volume, $C$ can be expressed in terms of the volume $V$ and the sinus of the dihedral angle $\theta_a$ (see equation (\ref{equa1})). To integrate this equation, we first express explicitly~\footnotemark it in terms of $l_a$, $l_b$ and $l_J$,
\footnotetext{$\f{\pd F^{(2)}}{\pd l_a}(l_j)=-\f{1}{2304\left((4l_J^2-l_b^2)l_a^3(4l_J^2-l_a^2)^3(4l_J^2-l_a^2+l_b^2)^4l_b^2\right)}(-1604l_a^8l_b^8l_J^2+1250816l_a^4l_J^8l_b^6-207104l_a^6l_b^6l_J^6+31904l_a^{10}l_J^4l_b^4
 -169344l_a^4l_b^8l_J^6-3920l_a^6l_b^{10}l_J^2+24992l_a^6l_b^8l_J^4-46848l_a^{10}l_J^6l_b^2-7129088l_a^4l_J^{10}l_b^4+1770496l_a^6l_J^8l_b^4+16832l_a^4l_b^{10}l_J^4
+34368l_a^8l_J^4l_b^6-6816l_a^{12}l_J^4l_b^2-278912l_a^8l_J^6l_b^4+14524416l_a^2l_J^{12}l_b^4+486144l_a^2l_J^8l_b^8-560l_b^{12}l_a^4l_J^2-43776l_a^2l_b^{10}l_J^6-3317760la^2lJ^{10}l_b^6
+22241280l_a^4l_J^{12}l_b^2+794l_a^{12}l_b^6-6955008l_a^6l_J^{10}l_b^2+2801664l_J^{12}l_b^6+672l_a^{14}l_b^2l_J^2-26542080l_a^4l_J^{14}-451584l_J^{10}l_b^8+46080l_J^8l_b^{10}
-2304l_b^{12}l_J^6-21233664l_J^{18}+37158912l_a^2l_J^{16}+1072128l_a^8l_J^8l_b^2-1528la^{12}l_b^4l_J^2-10911744l_J^{14}l_b^4-35979264l_a^2l_J^{14}l_b^2+1728l_b^{12}l_J^4l_a^2
+9953280l_a^6l_J^{12}+228096l_a^{10}l_J^8-10368l_a^{12}l_J^6-2073600l_a^8l_J^10+27l_a^{10}l_b^8+23592960l_J^{16}l_b^2+400l_a^8l_b^{10}-88l_a^{14}l_b^4-8144l_a^{10}l_b^6l_J^2+100l_b^{12}l_a^6)$}
and then deduce $F^{(2)}$:
\equa{ \tabl{ll}{ \label{NNLO}
F^{(2)}(l_j)&=\f{-1}{4608\left((4l_J^2-l_a^2)^2(4l_J^2-l_b^2)^2(4l_J^2-l_a^2-l_b^2)^3l_a^2l_b^2\right)}(-2359296l_a^2l_J^{10}l_b^4-224512l_a^6l_J^6l_b^4+100l_a^{12}l_b^4+576l_J^4l_b^{12}+112896l_J^8l_b^8\\
&+2727936l_a^4l_J^{12}+5308416l_J^{16}+212l_b^{10}l_a^6-5898240l_a^2l_J^{14}-11520l_a^{10}l_J^6+941056l_a^4l_J^8l_b^4+31584l_a^8l_J^4l_b^4\\
&-2416l_a^4l_b^{10}l_J^2
-79872l_a^8l_J^6l_b^2-480l_b^{12}l_J^2l_a^2-7040l_a^8l_b^6l_J^2-2416l_a^{10}l_J^2l_b^4+100l_a^4l_b^{12}+212l_a^{10}l_b^6\\
&+2727936l_J^{12}l_b^4-700416l_J^{10}l_b^6-5898240l_J^{14}
l_b^2-11520l_J^6l_b^{10}-700416l_a^6l_J^{10}+609l_a^8l_b^8+112896l_a^8l_J^8\\
&+576l_a^{12}l_J^4-2359296l_a^4l_J^{10}l_b^2+528384l_a^6l_J^8l_b^2+5849088l_a^2l_J^{12}l_b^2-79872l_a^2l_b^8l_J^6+31584l_a^4l_b^8l_J^4-7040l_a^6l_b^8l_J^2\\
&-224512l_a^4l_b^6l_J^6+58816l_a^6l_b^6l_J^4+8640l_a^{10}l_J^4l_b^2+8640l_b^{10}l_J^4l_a^2-480l_a^{12}l_J^2l_b^2+528384l_a^2l_J^8l_b^6)
}}
which is the only result with the required symmetries\footnote{If the result is not symmetric after integration, a non-null integration constant has to be added and  its determination can be done using the symmetry properties of the \sj-symbol. Indeed, we have $\f{\pd F^{(2)}}{\pd l_a}= H(l_a, l_b, l_J)$ so by integration over $l_a$, $F^{(2)}(l_j)=h(l_a, l_b, l_J)+Z(l_b, l_J)$. Moreover by symmetry, we must have $\f{\pd F^{(2)}}{\pd l_b}= H(l_a=l_b, l_b=l_a, l_J)$ and then integrating over $l_b$, we obtain a second expression for $F^{(2)}$: $F^{(2)}(l_j)=h(l_a=l_b, l_b=l_a, l_J)+Z(l_a, l_J)$ which implies that the constant of integration satisfies $Z(l_b,l_J)-Z(l_a,l_J)=h(l_a=l_b, l_b=l_a, l_J)- h(l_a, l_b, l_J)$. This equation allows to determine $Z$ and to get (\ref{NNLO}).}. The geometrical meaning of this function does not seem obvious. Nevertheless, we can give a more compact expression for the denominator of $F^{(2)}$:
\equa{\label{denoF2}
(4l_J^2-l_a^2)^2(4l_J^2-l_b^2)^2(4l_J^2-l_a^2-l_b^2)^3l_a^2l_b^2= \f{(12V)^6}{\cos^4\theta_J}.
}
\item The next equation comes from the terms of order $\lambda^{-3}$ which are proportional to $\cos(S_R+\f\pi4)$:
\equa{
\f{\pd G^{(2)}}{\pd l_a}(l_j)=0
}
which implies once again that $G^{(2)}=Z(l_b,l_J)$ is a constant of integration. Then the symmetry properties of the \sj-symbol implies $G^{(2)}(l_j)=0$.
\end{itemize}

\begin{figure}[ht]
\begin{center}
\includegraphics[width=4cm]{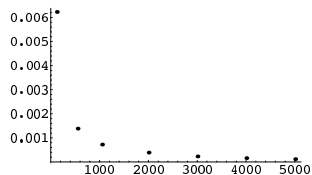}
\caption{Difference between the equilateral \sj-symbol and the analytical result (\ref{equaNNLO}). The x-axis stands for $l=d/2$ and $d$ goes from 100 to 5000. The error decreases as expected as $l^{-1}$ confirming our asymptotic formula.} \label{plotequiNNLO}
\end{center}
\end{figure}

We can now give the asymptotic expansion of an isosceles \sj-symbol until the next to next to leading order (NNLO):
\equa{ \label{isoNNLO}
\{ l_a, l_b\}^{\textrm{NNLO}}_{l_J}= \f{1}{\sqrt{12\pi V_{l_J}(l_a,l_a)}} \left[\cos (S_R +\f\pi4)+F^{(1)}(l_j) \sin(S_R+\f\pi4)+F^{(2)}(l_j) \cos(S_R+\f\pi4) \right]
}
where the expression for $F^{(1)}$ and $F^{(2)}$ are given by equations (\ref{geometricNLO}) and (\ref{NNLO}). This result seems to confirm that the expansion of the \sj-symbol is a series alternating cosines and sinus of the Regge action (shift by $\f\pi4$). In the case of an equilateral tetrahedron, all the edges have the same length, that is $l_a=l_b=l_J=l$ and $V= \f{\sqrt{2}}{12} l^3$. Then equation (\ref{isoNNLO}) reduces to:
\equa{\label{equaNNLO}
\{6j\}^{\textrm{NNLO}}_{\textrm{equi}}= \f{1}{\sqrt{\pi l^3\sqrt{2}}} \cos(S_R+\f\pi4)-\f{ 31}{72\,2^{1/4}\,2^{5/2}\sqrt{\pi l^5}}\sin(S_R +\f\pi4)-\f{45673}{20736}\f{1}{\,2^{1/4}\,2^{4}\sqrt{\pi l^7}}\cos(S_R+\f\pi4)
}
where the Regge action is given by $S_R=6 l \theta$ and $\theta=\theta_a=\theta_b=\theta_J= \arccos(-1/3)$.
This result is confirmed by numerical simulations. The plot figure \ref{plotequiNNLO} represents numerical simulations of the equilateral \sj-symbol minus its approximation (\ref{equaNNLO}). Moreover, to enhance the comparison, we have multiplied by $l^{7/2}$ to see how the coefficient of the NNLO is approached and we have divided by $\sin(S_R+\f\pi4)$ (oscillations of the next to next to next to leading order) to suppress the oscillations. This gives an error that decreases as expected as $l^{-1}$.

\begin{itemize}
\item The next two equations come from the terms of order $\lambda^{-4}$. The equation for $G^{(3)}$ is the same as the one for $G^{(1)}$ and $G^{(2)}$, that is: $\f{\pd G^{(3)}}{\pd l_a}=0$. Using the same arguments of symmetry we deduce that $G^{(3)}=0$. This confirms our expectation of a series alternating cosines and sines in the asymptotic of the \sj-symbol:
\equa{\label{asympt}
 \{\lambda l_a, \lambda l_b\}_{\lambda l_J}=\frac{1}{\lambda^{3/2}D(l_a,l_b,l_J)}\left[\cos(\lambda S_R+\f\pi4)+ \displaystyle{\sum_{k=1}^\infty}\frac{F^{(k)}(l_a,l_b,l_J)}{\lambda^k}\cos(\lambda S_R+\f\pi4+\eps(k)\f\pi2)\right],
}
where $\eps(k)=-1$ when $k$ is odd and $\eps(k)=0$ when $k$ is even.
We already have an expression for $F^{(1)}$ and $F^{(2)}$. The equation for $F^{(3)}$ is the second equation of order $\lambda^{-4}$ and gives its first derivative with respect to $l_a$ in terms of $l_a$, $l_b$ and $l_J$. It is straightforward (though lengthy) to integrate it over $l_a$ and get the expression\footnotemark of $F^{(3)}$ in terms of $l_a$, $l_b$ and $l_J$. In the equilateral case ($l_a=l_b=l_J=l$), the formula reduces to:
\equa{
F^{(3)}=\frac{28833535}{17915904}\frac{1}{2^{9/2}2^{1/4}l^3}
}
\footnotetext{$F^{(3)}(l_j)=-\f{1}{3317760(l_a^3(l_a-2l_J)(l_a+2l_J)(4l_J^2-l_b^2)^3(4l_J^2-l_a^2-l_b^2)^(9/2)l_b^3(4l_J^2-l_a^2)^2)}(-6965741568l_a^6l_J^{10}l_b^8+1069728l_a^6l_b^{16}l_J^2+28270080l_a^{16}l_J^6l_b^2-743040l_J^4l_a^2l_b^{18}+1750503260160l_J^{20}l_a^2l_b^2+379404800l_a^{12}l_J^6l_b^6+788705280l_a^{10}l_J^{10}l_b^4-33547200l_a^6
l_b^{14}l_J^4-98578608l_a^{12}l_b^8l_J^4-81032970240l_a^8l_J^{14}l_b^2+379404800l_a^6l_b^{12}l_J^6-389214720l_J^8l_a^2l_b^{14}-323813376000l_a^6l_J^{18}-156036l_a^8l_b^{16}-22176l_J^2l_a^4l_b^{18}-3824640l_a^{16}l_J^4l_b^4-28408l_a^{18}l_b^6-31104000l_a^{16}l_J^8-1262270545920l_J^{18}l_a^2l_b^4-1262270545920l_J^{18}l_a^4l_b^2+461814497280l_J^{16}l_a^2l_b^6+1069728l_a^{16}l_J^2l_b^6+74144841728l_a^6l_J^{12}l_b^6+824520278016l_J^{16}l_a^4l_b^4+461814497280l_a^6l_J^{16}l_b^2+788705280l_J^{10}l_a^4l_b^{10}-267541217280l_a^6l_J^{14}l_b^4+115174656l_J^6l_a^4l_b^{14}-999532800l_J^8l_a^4l_b^{12}+68611276800l_a^8l_J^{16}-3824640l_J^4l_a^4l_b^{16}-1242078720l_a^{10}l_J^8l_b^6+28270080l_J^6l_a^2l_b^{16}+38483472384l_J^{12}l_a^4l_b^8+2157096960l_a^{12}l_J^{10}l_b^2-1242078720l_a^6l_b^{10}l_J^8+509607936000l_J^{24}-33547200l_a^{14}l_b^6l_J^4+1222041600l_J^{12}l_a^2l_b^{10}+11119152l_a^{12}l_b^{10}l_J^2-999532800l_a^{12}l_J^8l_b^4+783522201600l_J^{20}l_a^4+2157096960l_J^{10}l_a^2l_b^{12}-28408l_a^6l_b^{18}-389214720l_a^{14}l_J^8l_b^2-783601920l_a^8l_J^8l_b^8-98578608l_a^8l_J^4l_b^{12}-6965741568l_a^8l_J^{10}l_b^6-323813376000l_J^{18}l_b^6+1222041600l_a^{10}l_J^{12}l_b^2+115174656l_a^{14}l_J^6l_b^4-743040l_a^{18}l_J^4l_b^2-22176l_a^{18}l_J^2l_b^4-81032970240l_J^{14}l_a^2l_b^8+7201368l_a^{14}l_b^8l_J^2-114011616l_a^{10}l_b^{10}l_J^4-985242009600l_J^{22}l_b^2+568565760l_a^{10}l_b^8l_J^6-1401753600l_a^{12}l_J^{12}-267541217280l_J^{14}l_a^4l_b^6+568565760l_a^8l_J^6l_b^{10}+11119152l_a^{10}l_b^{12}l_J^2+783522201600l_J^20l_b^4-342523l_a^{12}l_b^{12}+38483472384l_a^8l_J^{12}l_b^4-3981312000l_a^{10}l_J^{14}-156036l_a^{16}l_b^8-511602l_a^{10}l_b^{14}+344217600l_b^{14}l_J^{10}-1401753600l_b^{12}l_J^{12}+1036800l_b^{18}l_J^6+344217600l_a^{14}l_J^{10}+68611276800l_J^{16}l_b^8-3981312000l_J^{14}l_b^{10}-511602l_a^{14}l_b^{10}-31104000l_b^{16}l_J^8-985242009600l_J^{22}l_a^2+1036800l_a^{18}l_J^6+7201368l_a^8l_b^{14}l_J^2)$}
Therefore, we have the expression of the asymptotic expansion of the equilateral \sj-symbol up to the next-to-next-to-next to leading order (NNNLO):
\bes
\{6j\}^{\textrm{NNNLO}}_{\textrm{equi}}&= \f{1}{2^{1/4}\,\sqrt{\pi l^3}} &
\left[ \cos(S_R+\f\pi4)-\f{31}{72\,2^{5/2}\,l}\sin(S_R +\f\pi4)\right.\nn\\
&&-\left.\f{45673}{20736\,2^4\,l^2}\cos(S_R+\f\pi4) +\frac{28833535}{17915904\,2^{9/2}\,l^3}\sin(S_R+\f\pi4)\right].
\ees
We check this result numerically by computing it using Mathematica for two values of spins $j=50$ and $j=100$. More precisely, we computed the renormalized error $\f{(\{6j\}_{\textrm{equi}}-\{6j\}^{\textrm{NNNLO}}_{\textrm{equi}})l^{9/2}}{\cos(S_R+\f\pi4)}$ and we got the expected $1/\lambda $-behavior. However for $l> 100$, Mathematica is not accurate enough and the numerical errors are too important to get exploitable results.
In the general isosceles case, the expression of $F^{(3)}$ is quite complicated and its geometrical interpretation remains to be understood. Nevertheless, we can again give as before a more compact formula for the denominator of $F^{(3)}$:
\equa{\label{denoF3}
\textrm{denominator}_{F^{(3)}}=3317760 (4l_J^2-l_a^2-l_b^2)^{9/2}l_a^3l_b^3(4l_j^2-l_a^2)^3(4l_J^2-l_b^2)^3= 30(48)^3\f{(12V_J(a,b))^9}{\cos^6\theta_J}
}
From this equation, the equation giving the denominator of $F^{(2)}$ and remembering that the denominator of $F^{(1)}$ can be written under a similar form: denominator$_{F^{(1)}}=48\frac{(12V)^3}{\cos^2\theta_J}$, we can conjecture that:
\equa{
\textrm{denominator}_{F^{(k)}} \propto \frac{(12V_J(a,b))^{3k}}{(\cos \theta_J)^{2k}}
}
where $F^{(k)}$ are the terms appearing in the asymptotic expansion of the \sj-symbol (\ref{asympt}). And consequently, the numerator of $F^{(k)}$ is a polynomial in $l_j$ of degree $8k$.
\end{itemize}
So, using the recursion relation for the isosceles \sj-symbol as well as its symmetry properties, we have computed explicitly the asymptotic expansion of the isosceles \sj-symbol to the fourth order up to an overall factor $K$ (this integration constant $K$ comes from the integration of the first equation (\ref{equa1})). The well-known value $K=\sqrt{12 \pi}$ which already appears in the Ponzano-Regge formula  can be obtained easily using the unitary property of the \sj-symbol, as we show in the next section.  The equilateral case has been checked against numerical calculations. This method using the recursion relation is fairly easy to implement. It requires integrating a rational fraction at each level and does not involve neither Riemann sum nor saddle point analysis.
Moreover, since the coefficient $C$ of the recursion relation (\ref{exactrecursion}) is a polynomial of degree 3; $\f{\pd^n C}{\pd l_a^n}=0$ for $n \geq 4$. Therefore, we expect to get a stable relation for the first derivative of $F^{(k)}$ and $G^{(k)}$ with respect to $l_a$ for  $k \geq 3$. On one hand, this allows to prove that $G^{(k)}$ always vanishes; and on the other hand, it should provide a systematic method to extract $F^{(k)}$ for arbitrary order $k$.


We conclude this section with a general remark on the asymptotic expansion of the \sj-symbol. In the context of 3d quantum gravity, it is often argued that the leading order of the \sj-symbol is a $\cos(S)$ instead of a complex phase $\exp(+iS)$, thus reflecting that the path integral is invariant under a change of (local) orientation (see e.g.\cite{laurent}). This obviously neglects the $+\pi/4$ shifts, which can be considered as a quantum effect (like an ordering ambiguity). However, in the light of the present expansion, it is clear that we have terms of the type $\sin(S)$ beyond the leading order and such terms are not invariant under the change $S\arr -S$. This means that the role of this symmetry in the spinfoam path integral should be more subtle than originally thought.

\subsection{Consequences of the unitary property of the \sj-symbols}

The orthogonality property of the \sj-symbols states that:
\equa{
\sum_{l_a} 4l_a \sqrt{l_b l_{b^{\prime}}} \{l_a, l_b\}_{l_J} \{l_a, l_{b^{\prime}}\}_{l_J}=\delta_{l_b l_{b^{\prime}}}
}
This relation corresponds to the unitarity of the evolution in the Ponzano-Regge 3d quantum gravity.
We want to use this property to determine the constant of the leading order of the \sj-symbol. From the recursion relation we have shown that $\{l_a, l_b\}^{\textrm{LO}}_{l_J}= \f{K}{\sqrt{V_J(a,b)}} \cos(S_R+\f\pi4)$; but $K$ is still undetermined. For large spin and for $l_b \approx l_{b^\prime}$, we can approximate the unitary property at the leading order in $(l_b-l_{b^\prime})$ by:
\equa{\label{unitary}
 \int_0^\infty dl_a 4 l_a l_b \f{K^2}{V_J(a,b)}\cos(S_R(l_a, l_b)+\f\pi4) \cos(S_R(l_a, l_{b^\prime})+\f\pi4)\approx \delta(l_b-l_{b^\prime } ).
}
The product of the cosines can be simplified at leading order:
\bes
\cos(S_R(l_a, l_b)+\f\pi4) \cos(S_R(l_a, l_{b^\prime})+\f\pi4)
&=&
\f12\left[\cos(S_R(l_a, l_b)+S_R(l_a, l_{b^\prime})+\f\pi2)+\cos(S_R(l_a, l_b)-S_R(l_a, l_{b^\prime}))\right] \nn\\
&\sim&
\f12\left[\cos(2S_R(l_a, l_b)+\f\pi2)+\cos((l_b-l_{b^\prime})\theta_b)\right], \nn
\ees
%
where the dihedral angle $\theta_b=\arccos \left( -\f{4l_J^2-l_b^2-2l_a^2}{4l_J^2-l_b^2}\right)$ is considered as a function of the length $l_a$. We do a saddle point approximation. The first term oscillates and its integral is exponentially suppressed. And we are left with the second term, which should satisfy the following equation:
\equa{
 \int_{-\infty}^\infty dl_a  l_a l_b \f{K^2}{V_J(a,b)}\cos((l_b-l_{b^\prime }) \theta_b) \approx \delta(l_b-l_{b^\prime } )
}
We recall that:
$$
\f{1}{2\pi} \int_{-\infty}^\infty dl_a \cos(l_a(l_b-l_{b^\prime }))= \delta(l_b-l_{b^\prime })
$$
therefore we can conclude that
\equa{
l_al_b \f{K^2}{V}=\f{1}{2\pi} \left| \f{\pd \theta_b}{\pd l_a} \right|.
}
$\theta_b$ and $l_b$ are so conjugate variables and $K$ comes from the Jacobian of the change of variables between $l_a$ and $\theta_b$.
Computing the derivative of the dihedral angle gives:
\be
\f{\pd \theta_b}{\pd l_a}= \f{-2}{\sqrt{4l_J^2-l_a^2-l_b^2}}=\f{-l_al_b}{6V_J(a,b)}
\quad
\Rightarrow
\quad
K=\f{1}{\sqrt{12\pi}}.
\ee
Moreover, pushing the approximation of the unitary property to the next to leading order in $(l_b-l_{b^\prime})$ and using the next to leading order of the \sj-symbol shows that $G^{(1)}=0$. This was already shown in the previous part using the recursion relation and the symmetry properties of the \sj-symbol and comes as a confirmation.

\section{``Ward-Takahashi identities" for the spinfoam graviton propagator}

We are interested in the two-point function in 3d quantum gravity for the simplest triangulation given by a single tetrahedron. This provides the first order of the ``spinfoam graviton propagator" in 3d quantum gravity.

Considering the isosceles tetrahedron, we focus on the correlations between the two representations $a$ and $b$:
\be
\la \cO(a) \tcO(b)\ra_{\psi_J}=
\f{1}{Z} \,\sum_{a,b}\psi_J(a)\psi_J(b)\cO(a) \tcO(b) \{a,b\}_J,
\qquad
Z\equiv\,\sum_{a,b}\psi_J(a)\psi_J(b)\{a,b\}_J,
\ee
where $\psi_J(j)$ is the boundary state, which depends also on the bulk length scale $J$, and $\cO,\tcO$ are the observables whose correlation we are studying.

Now, inserting a recursion relation with shifts on $a$, $b$ or $J$ in the sum over the representation labels $\sum_{a,b}$ leads to equations relating the expectation values of different observables. We distinguish two cases: when the state $\psi_J$ does not change or when the length scale $J$ also varies.

\subsection{Relating Observables}

Inserting the recursion relation on $a$-shifts in the definition of the correlation function, we obtain the following exact identity:
\be\tabl{ll}{
\la \f{\psi_J(a-1)}{\psi_J(a)}\cO(a-1)\tcO(b)(l_a-\f12)(4l_J^2-(l_a-\f12)^2)\ra_\psi&-
\la \cO(a)\tcO(b)2l_a(2\cos \theta_a (4l_J^2-l_a^2)+\f14)\ra_\psi\\
&+ \la \f{\psi_J(a+1)}{\psi_J(a)}\cO(a+1)\tcO(b)(l_a+\f12)(4l_J^2-(l_a+\f12)^2)\ra_\psi
=0.
}\ee
We call this a Ward identity for our spinfoam correlation.
If the observable diverges at $a=0$, more precisely if it contains terms in $1/a$ or in $1/(a+1)$, then we need to take into account extra boundary terms in this equation corresponding to contributions at $a=0$. But all observables usually considered are regular in this sense.

Then one can choose different sets of observables $\cO$ and $\tcO$ and one gets different identities on the correlation functions of the spinfoam model. 
 For example, taking $\cO(a)=l_a$, we get:
$$ \tabl{ll}{
\la \f{\psi_J(a-1)}{\psi_J(a)}\tcO(b)(l_a-1)(l_a-1/2)(4l_J^2-(l_a-1/2)^2)\ra_\psi&-
\la \tcO(b)(2\cos \theta_a l_a^2(4l_J^2-l_a^2)+l_a^2/2)\ra_\psi\\
&+\la \f{\psi_J(a+1)}{\psi_J(a)}\tcO(b)(l_a+1)(l_a+1/2)(4l_J^2-(l_a+1/2)^2)\ra_\psi
=0.
}$$
We recall that the area of the triangle of edge lengths given by $l_a$, $l_J$, $l_J$ is equal to $A(l_a, l_J)= \f14 l_a\sqrt{4l_J^2-l_a^2}$; then $(l_a\pm1)(l_a \pm 1/2)(4l_J^2-(l_a\pm 1/2)^2)=16[A^2(l_a\pm1/2,l_J)\pm \f{A^2(l_a\pm1/2,l_J)}{2(l_a\pm 1/2)}]$, therefore we can rewrite the previous equation as  an equation between correlation functions of the observable $\tcO(b)$ and different observables proportional to the square of the triangle area $A(l_a, l_J)$:
$$ \tabl{ll}{
\la \f{\psi_J(a-1)}{\psi_J(a)}[A^2(l_a-1/2,l_J)-\f{A^2(l_a-1/2,l_J)}{2(l_a- 1/2)}]\tcO(b)\ra_\psi&-
\la(2\cos \theta_aA^2(l_a,l_J)+l_a^2/2)\tcO(b)\ra_\psi\\
&+\la \f{\psi_J(a+1)}{\psi_J(a)}[A^2(l_a+1/2,l_J)+\f{A^2(l_a+1/2,l_J)}{2(l_a+ 1/2)}]\tcO(b)\ra_\psi
=0.
}$$
The standard choice of boundary is a phased Gaussian \cite{graviton, 3dtoymodel, physical}:
\be
\psi_J(j) \,\sim\, e^{i2l_j\vtheta} e^{-2\alpha\f{(l_j-l_J)^2}{l_J}},
\ee
where $\vtheta$ is a fixed angle defining a posteriori the external curvature of the boundary and $\alpha$ is an arbitrary real positive number (which can be fixed by the requirement of a physical state \cite{physical}).
In this case, we can compute explicitly the ratios $\psi(a\pm1)/\psi(a)$ entering the Ward identity:
$$
\f{\psi_J(a\pm 1)}{\psi_J(a)}= e^{\pm i2\vtheta} e^{\mp 4 \alpha \f{l_a-l_J}{l_J}} e^{-\f{2\alpha}{l_J}}
$$
Of course, this ratios does not depend on $b$; therefore if the observable $\tcO(b)=1$, then the dependence on $b$ only appears in one correlation function through the cosine of the dihedral angle $\theta_a$.
%
As another example, we consider $\cO(a)=l_a^{-1}$ and $\tcO(b)=\f{4l_J^2-l_b^2}{(2l_J)^{4}}$, then:
$$\tabl{ll}{
\la \f{\psi_J(a-1)}{\psi_J(a)}\, \f{l_a-1/2}{l_a-1}\,\f{4l_J^2-(l_a-1/2)^2}{4l_J^2}\, \f{4l_J^2-l_b^2}{4l_J^2} \ra_\psi &-2\la \cos\theta_a\f{4l_J^2-l_a^2}{4l_J^2}\, \f{4l_J^2-l_b^2}{4l_J^2} + \f{1}{16l_J^2}\, \f{4l_J^2-l_b^2}{4l_J^2} \ra_\psi \\ &+ \la \f{\psi_J(a+1)}{\psi_J(l_a)} \,\f{l_a+1/2}{l_a+1}\, \f{(4l_J^2-(l_a+1/2)^2}{4l_J^2}\, \f{4l_J^2-l_b^2}{4l_J^2}\ra_\psi=0
}$$
which can be approximated by:
$$
\la e^{- i2\vtheta} e^{ 4 \alpha \f{l_a-l_J}{l_J}}\,\Delta((l_a-1/2)^2)\Delta(l_b^2) \ra_\psi -2e^{\f{2\alpha}{l_J}}
\la \cos\theta_a\Delta(l_a^2)\Delta(l_b^2)+ \f{1}{16l_J^2}\Delta(l_b^2) \ra_\psi + \la e^{ i2\vtheta} e^{-4 \alpha \f{l_a-l_J}{l_J}} \,\Delta((l_a+1/2)^2)\Delta(l_b^2)\ra_\psi \approx 0
$$
where $\Delta(l_j^2)=\f{l_j^2-4l_J^2}{4l_J^2}$.

\subsection{Rescaling the Tetrahedron}

We can now vary also the length scale $l_J$.  First let's notice that in the same way we wrote an exact recursion relation for the leading order of the isosceles  \sj-symbol shifting the representation $a$ (equation (\ref{exactrecurLO})), we can write a similar exact recursion relation for the leading order of the \sj-symbol shifting the label $J$; that is
\equa{
\sqrt{V_{J+1}(a,b)} \{a,b\}^{\textrm{LO}}_{J+1}-2\cos(4\theta_J)\sqrt{V_{J}(a,b)} \{a,b\}^{\textrm{LO}}_{J}+\sqrt{V_{J-1}(a,b)} \{a,b\}^{\textrm{LO}}_{J-1}=0
}
Inserting this recursion relation on $J-$shifts in the definition correlation function, we obtain the following identity:
\bes
\la \sqrt{V_{J+1}(a,b)} \f{\psi_J(a)\psi_J(b)}{\psi_{J+1}(a)\psi_{J+1}(b)} \cO(a) \tcO(b) \ra_\psi
&+\la \sqrt{V_{J-1}(a,b)} \f{\psi_J(a)\psi_J(b)}{\psi_{J-1}(a)\psi_{J-1}(b)} \cO(a) \tcO(b) \ra_\psi&\nn\\
&-2\la \cos(4\theta_J) \sqrt{V_{J}(a,b)} \cO(a) \tcO(b) \ra_\psi &
=0
\ees
The correlation functions appearing in this equation are in fact approximation. We are allowed to use the leading order of the \sj-symbol because the boundary state used picks the function on large $j_0$. And for the same reason, we can expand $\sqrt{V_{J\pm1}(a,b)}$ and the ratios $\f{\psi_J(a)\psi_J(b)}{\psi_{J\pm1}(a)\psi_{J\pm1}(b)}$:
\equa{\tabl{l}{
\la \sqrt{V_{J}(a,b)}\left(1-\f{2l_J}{4l_J^2-l_a^2-l_b^2}\right) e^{-4\alpha\f{(2l_J-(l_a+l_b))}{l_J}[1+\f{3l_J-2(l_a+l_b)}{2l_J(2l_J-l_a-l_b)}]} \cO(a) \tcO(b) \ra_\psi  -2\la \cos(4\theta_J) \sqrt{V_{J}(a,b)} \cO(a) \tcO(b) \ra_\psi \\
\quad \quad \quad \quad \quad \quad+\la \sqrt{V_{J}(a,b)}\left(1+\f{2l_J}{4l_J^2-l_a^2-l_b^2}\right) e^{4\alpha\f{(2l_J-(l_a+l_b))}{l_J}[1-\f{3l_J-2(l_a+l_b)}{2l_J(2l_J-l_a-l_b)}]} \cO(a) \tcO(b) \ra_\psi \approx 0.
}}
We hope that such equation will turn out useful to study the asymptotic properties of the correlations function as the length scale $J$ grows large, but we leave this for future investigation.

\section*{Conclusion}

We have used the recursion relation satisfied by the \sj-symbol to study the structure of its asymptotical expansion for large spins. The exact recursion relation allowed us to compute explicit the asymptotical approximation of the isosceles \sj-symbol up to fourth order. This confirms previous results \cite{valentin,maite} and introduces techniques allowing further systematic analytical calculations of the corrections to the behavior of the\sj-symbol at large spins. However a clear and simple geometrical interpretation of the polynomials appearing in this expansion is still missing, but the differential equations that we provide for these coefficients should be a first step in this direction.

This work is useful in particular for the study of large scale correlations in the spinfoam model for 3d quantum gravity. In this context, the recursion relation allowed us to write equations satisfied by the spinfoam correlations similar to the Ward identities of standard quantum field theory. We hope that such recursion techniques can be further applied to the study of 4d spinfoam amplitudes and the resulting spinfoam graviton propagator \cite{recursion}.

\section*{Ackowledgements}

The numerical simulations and plots were done using Mathematica 5.0.
MD and ER are partially supported by the ANR ``Programme Blanc" grant LQG-06.




\end{document}